\documentclass[10pt,letterpaper,aps,pra,twocolumn,superscriptaddress]{revtex4-1}
\usepackage[latin1]{inputenc}
\usepackage{amsmath}
\usepackage{amsfonts}
\usepackage{amssymb}
\usepackage{bm}
\usepackage[pdftex]{graphicx}
\usepackage[colorinlistoftodos]{todonotes}

\begin{document}
\title{Power-recycled weak-value-based metrology}
\author{Kevin Lyons}
\affiliation{Department of Physics and Astronomy, University of Rochester, Rochester, New York 14627, USA}
\affiliation{Center for Coherence and Quantum Optics, University of Rochester, Rochester, New York 14627, USA}
\author{Justin Dressel}
\affiliation{Department of Electrical and Computer Engineering;	University of California, Riverside, California 92521, USA}
\author{Andrew N. Jordan}
\affiliation{Department of Physics and Astronomy, University of Rochester, Rochester, New York 14627, USA}
\affiliation{Center for Coherence and Quantum Optics, University of Rochester, Rochester, New York 14627, USA}
\affiliation{Institute for Quantum Studies, Chapman University, 1 University Drive, Orange, CA 92866, USA}
\author{John C. Howell}
\affiliation{Department of Physics and Astronomy, University of Rochester, Rochester, New York 14627, USA}
\affiliation{Center for Coherence and Quantum Optics, University of Rochester, Rochester, New York 14627, USA}
\author{Paul G. Kwiat}
\affiliation{Department of Physics, University of Illinois at Urbana-Champaign, Urbana, Illinois 61801-3080, USA}
\date{\today}
\begin{abstract}
We improve the precision of the interferometric weak-value-based beam deflection measurement by introducing a power recycling mirror, creating a resonant cavity.  This results in \emph{all} the light exiting to the detector with a large deflection, thus eliminating the inefficiency of the rare postselection.  The signal-to-noise ratio of the deflection is itself magnified by the weak value.  We discuss ways to realize this proposal, using a transverse beam filter and different cavity designs.
\end{abstract}

\newcommand{\op}[1]{\hat{ #1}}                
\newcommand{\ket}[1]{\lvert#1\rangle}
\newcommand{\bra}[1]{\langle#1\rvert}
\newcommand{\pr}[1]{\ket{#1}\bra{#1}}
\newcommand{\ipr}[2]{\langle #1 | #2 \rangle}
\newcommand{\mean}[1]{\left\langle #1 \right\rangle}
\newcommand{\cw}{\circlearrowright}
\newcommand{\ccw}{\circlearrowleft}

\maketitle
The weak value amplification effect, introduced by Aharonov, Albert, and Vaidman \cite{Aharonov1988}, permits a small change in a system/meter coupling parameter to be converted into a large change in a meter variable.  This effect comes at the sacrifice of only measuring a small postselected fraction of the events experiencing the amplified meter variable.  This gain and loss balance each other, leading to the same signal-to-noise ratio (SNR) of the measured parameter as would be found if the measurement were made directly, provided the system is ideal \cite{Starling2010}.  The effect may also be viewed as a concentration of the Fisher information about the measured parameter into a small number of collected events \cite{Jordan2014,Viza2014,Pang2014}.  Combined with the fact that the weak values-based approach can perform better than the standard method in the presence of certain technical limitations, such as beam jitter noise or detector saturation \cite{Jordan2014,Viza2014}, ultra-sensitive optical beam displacement and deflection measurements have been achieved using this technique, see e.g. \cite{Hosten2008,Dixon2009,Hogan2011}.  For recent reviews of this and related topics, see Refs.~\cite{Kofman2011,Dressel2014}.  Here we focus on the interferometric weak value setup used in Ref.~\cite{Dixon2009}, which couples the transverse beam position to the ``which-path'' counter-propagating modes of a Sagnac interferometer.  The postselection corresponds to measuring only the light emerging from the ``dark'' port of the interferometer.  The meter change corresponds to a transverse beam deflection.

\begin{figure}[htp]
\includegraphics[width=\columnwidth]{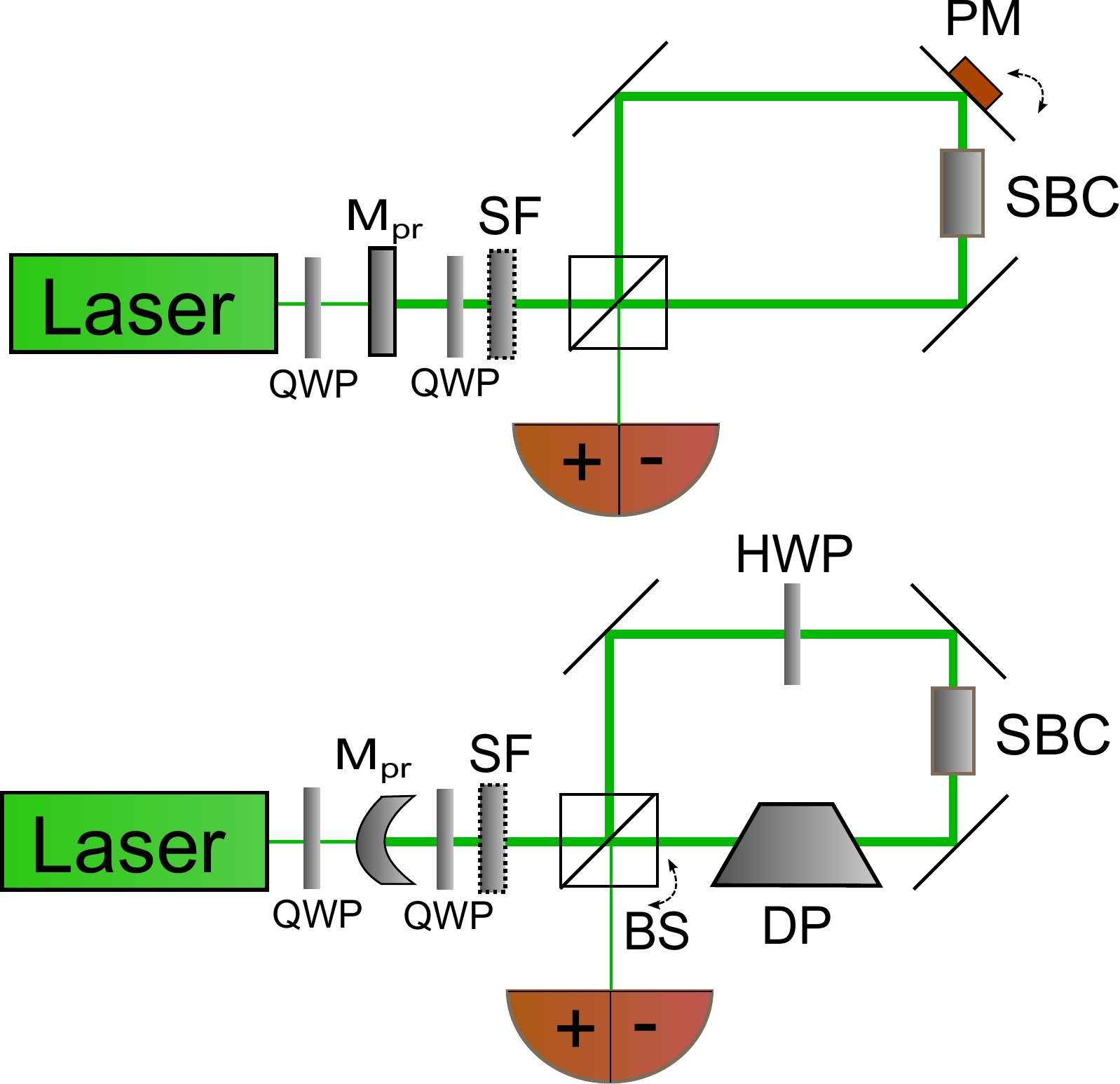}
\caption{Schematics of weak value-based metrology with power recycling cavities.  (a) Plane parallel cavity.  A partially reflecting mirror (M$_\text{pr}$), forms an optical cavity in combination with a Sagnac interferometer, causing all the light to exit to the detector.  The tunable phase difference between the counter-propagating arms is controlled by a Soleil-Babinet Compensator (SBC). A piezo-driven mirror (PM) imparts a small deflection to the beam, that is measured at a split detector. A spatial filter (SF) refreshes the beam profile on each pass.  (b) Confocal cavity.  A symmetric confocal resonator is established with the curved recycling mirror.  Here the 50/50 beamsplitter (BS) is tilted instead of the mirror (which is now at the focus).  A Dove prism (DP) inside the interferometer corrects for the profile flip when the Gaussian beam passes through its focus. The quarter waveplates (QWP) together with the SBC ensure the polarization optics leads to the same displacement at the detector on every pass.}
\label{figsetup}
\end{figure}

Unlike traditional deflection measurements that directly focus the beam onto the detector, the weak value deflection effect relies on wave interference.  As such, we can consider combining this effect with other interferometric techniques that have been useful for precision measurement schemes, such as in gravitational wave detectors. One such technique is power recycling, initially proposed by Drever \cite{Drever1983}.  By placing a partially transmitting mirror at the bright port of an interferometer to form a resonant cavity, one is able to more efficiently use the input laser power by increasing the total power inside the interferometer.  In the context of these weak value-based experiments, the photons that would have previously exited the interferometer through the bright port now destructively interfere, effectively trapping them inside the interferometer , so that, in the absence of loss, the entire input beam eventually exits the (formerly) dark port of the interferometer.

With the weak value deflection measurement inside the cavity, the full beam experiences the amplified deflection, which can then be detected with a position sensitive detector, such as a split detector.  Since the single-pass, postselected, weak value measurement displays the same SNR as the ideal focusing deflection technique \cite{Starling2010,Jordan2014,Viza2014}, the power-recycling enhancement permits the amplification of the SNR itself by the large weak value factor, thus surpassing both techniques.  We focus in this paper on the continuous wave recycling of power to increase the precision of the weak value technique. A related pulsed recycling scheme, using a Pockels cell to trap a pulse inside the interferometer, was proposed in Ref.~\cite{Dressel2013}.  One important difference between the two ideas is that the pulsed scheme relies on the Pockels cell and polarization optics to trap the pulse in the interferometer, whereas here, it is the destructive interference of the reflected field that causes all the light to exit to the detector.

{\it Weak value amplification.}---We first briefly review the interferometic weak value setup of Ref.~\cite{Dixon2009} using a Sagnac interferometer.  A phase difference $\phi$ between the counter-propagating modes of the interferometer may be controlled with a Soleil-Babinet Compensator (SBC) to break the clockwise/counterclockwise symmetry. A piezo-driven deflecting mirror is placed at the symmetry point of the interferometer to induce a small transverse momentum kick $k$, the parameter we wish to precisely measure. This parameter can be measured using a split detector placed at the output dark port.  

We consider a continuous wave laser with a transverse Gaussian profile $E_0(x) = (N^2/2\pi\sigma^2)^{1/4}\exp(-x^2/4\sigma^2)$ that we have normalized to the average number of available photons $N$ per unit time.  The split detector gives a signal as the difference between the number of photons on the left versus right side of the detector, $S = N_R - N_L$, per unit time.
The signal of the split detector at the dark port is linear in $k$ for the interferometric weak value measurement (assuming $k\sigma \ll \phi/2 \ll 1$) \cite{Dixon2009,Dressel2013},
\begin{align}\label{eq:signal}
  \mean{S} &\approx 2\sqrt{\frac{2}{\pi}}\,N_{\rm det}\,\frac{2k\sigma}{\phi},
\end{align}
where $N_{\rm det}$ is the total number of photons that are detected out of the initial beam.  From a quantum mechanical perspective, we note that the large factor of $2/\phi$ is related to the weak value of the which-path operator, $\op{W}$, given by $\langle \psi_f | \op{W} | \psi_i \rangle /\langle \psi_f | \psi_i \rangle  = -i \cot(\phi/2) \approx -2i/\phi$.  Here, $\op{W}=\pr{\cw}-\pr{\ccw}$ is defined with the two orthogonal circulating states $\ket{\cw}$ and $\ket{\ccw}$ of the Sagnac interferometer, and $|\psi_i\rangle$ and $| \psi_f \rangle$ are the pre- and post-selected states of the interferometer, defined by the phase $\phi$ and the selection of the port to measure.  The weak value effectively amplifies the kick $k$ for each collected photon.  The variance of this signal is limited by the detected shot noise $N_{\rm det}$, so the SNR ${\cal R}$ is given by
\begin{align}\label{eq:snr}
  {\cal R} &\equiv \frac{\mean{S}}{\sqrt{N_{\rm det}}} \approx 2\sqrt{\frac{2}{\pi}}\,\sqrt{N_{\rm det}}\,\frac{2k\sigma}{\phi}.
\end{align}

{\it Continuous-wave power recycling.}---To introduce power recycling, we modify the previous setup of Ref.~\cite{Dixon2009} by adding a partially transmitting mirror, illustrated in Fig.~\ref{figsetup}(a), to make the Sagnac interferometer a resonant optical cavity.  In the absence of the cavity, the detected number of photons $N_{\rm det} = p\,N$ is small when the postselection probability $p \approx (\phi/2)^2$ of the dark port is also small.  The resulting factor of $\sqrt{p}$ in Eq.~\eqref{eq:snr} thus exactly cancels the amplification of the large weak value factor, and precisely recovers the SNR that one expects from a traditional beam deflection measurement with all the light \cite{Starling2009}, but still with enhanced robustness to some types of technical noise \cite{Jordan2014,Viza2014}.  However, the addition of the cavity will permit the entire input beam to exit the (formerly) dark port and be detected with the amplified deflection.  As we will soon show, including the cavity ideally makes $N_{\rm det} = N$, so the large weak value $(2/\phi)$ directly amplifies the SNR itself in Eq.~\eqref{eq:snr}, giving a large prefactor to the standard quantum limit scaling.  We note this variation also turns the probabilistic weak value method into a deterministic one with respect to the output port.

To see why such a resonant cavity will permit the entire beam to be detected, consider an initial amplitude $E_0$ that is incident on a cavity formed by two partially transmitting mirrors with transmission, $t_1$ and $t_2$, and reflection, $r_1 = \sqrt{1-t_1}$ and $r_2 = \sqrt{1-t_2}$, coefficients.  For light inside the cavity, each round trip adds a phase $\theta$, which depends on the geometry of the cavity and the wavelength of light, giving a geometric series for the amplitude,
\begin{align}\label{eq:simplecavityamplitude}
\nonumber E_{\rm cav}&= t_1[1 + r_1 r_2 e^{i\theta} + (r_1 r_2 e^{i\theta})^2 + \cdots]\,E_0, \\
&=\frac{t_1}{1 - r_1 r_2 e^{i\theta}}\,E_0.
\end{align}   
The light reflected back towards the laser is similarly geometric, with the amplitude $E_r$ being a superposition of amplitudes from light directly reflected from the first mirror and multiple reflections inside the cavity,
\begin{align}\label{eq:simplecavityreflection}
E_{r} &= \left[-r_1 + \frac{t_1^2 r_2 e^{i\theta}}{1 - r_1 r_2 e^{i\theta}}\right]\,E_0.
\end{align}
If $\theta = 2\pi n$, where $n$ is any integer, and $r_1 = r_2 \equiv r$, then the reflected amplitude is exactly zero, so the power leaving the cavity through the second mirror becomes equal to the input laser power.  This condition is known as impedance matching \cite{Schnier1997}.  In this case the light intensity inside the cavity is amplified above the input laser intensity by a gain factor 
\begin{align}\label{eq:gaingeneral}
 G &\equiv \frac{|E_{\rm cav}|^2}{|E_0|^2} = \frac{t^2}{(1 - r^2)^2} = \frac{1}{t^2} = \frac{1}{T},
\end{align}   
equivalent to the inverse transmission probability $T$.  

We will soon show that these general results may be applied to our weak-value amplification cavity formed from one partially transmitting mirror and the partially transmitting dark port of the interferometer. We effectively replace the transmission $T$ in \eqref{eq:gaingeneral} by the postselection probability of exiting the interferometer, $T \rightarrow p \approx (\phi/2)^2$.  Thus, the large power gain inside the inteferometer allows for the small postselection probability to give the entire input beam out of one port, and no light out of the other port, boosting the SNR of Eq.~\eqref{eq:snr} by $1/\sqrt{p}$.

{\it Resource counting.---}The relative advantages of one technique to another should specify the resources given as a constraint.  We note that if the resource is taken to be the number of {\it detected} photons, then there is already an advantage in the single-pass weak value experiment over the direct deflection experiment (not considering technical noise sources) \cite{Dixon2009, Starling2009}.  If instead the resource is the total number of photons entering the interferometer, the power-recycled proposal gives an advantage $1/\sqrt{p}$ over the single pass experiment for many cycles. 
Perhaps a fairer way of counting resources is the number of times an interaction takes place.  In this proposal, the enhancement in sensitivity is due to an effective power increase, but we stress that signal does not accumulate for multiple passes.  The total number of interactions with the unknown parameter is $N M$, where $N$ is the total photon number, and $M$ is the number of times a photon enters the interferometer.  While $M$ is a stochastic variable in our system, on average it is $1/p$, giving $N/p$ as the total number of interactions, whose square-root will be shown to determine ${\cal R}$ (\ref{eq:snrresult}).
Another interesting technique using multiple interactions where the signal scales linearly with the number of cycles,
while keeping the noise constant is a simple example of {\it signal recycling} \cite{Meers1991}.  In this case recycled photons accumulate additional momentum kicks on each traversal thereby enhancing the signal. This separate technique can be used together with power recycling as a complimentary method, and was already incorporated in this kind of measurement via an optical lever by the Kasevich group \cite{Hogan2011}.

{\it Recycling with flat mirrors.}---The remaining requirement for the weak value technique to still work in the presence of the cavity is for the transverse position profile at the beamsplitter to be preserved, so that the enhanced deflection remains with each traversal through the interferometer.  We now calculate the transverse profile for a beam confined by flat cavity mirrors by adapting the operator approach used in \cite{Dressel2013}.  While all intensities in this setup may be calculated from classical wave optics, it is convenient to adopt a quantum state analysis. The flat mirror approach is a reasonable approximation if the roundtrip distance of the cavity multiplied by the finesse is much less than the Rayleigh length of the beam. Later we will discuss a more realistic cavity design that uses a curved mirror to confine a Gaussian beam.

To determine the steady state beam profile at the detector, we introduce the ``system'' state $\ket{\psi}$ spanned by the orthogonal circulating modes $\ket{\cw}$ and $\ket{\ccw}$ of the Sagnac interferometer, and the ``meter'' state $\ket{\varphi}$ which represents the transverse profile of the beam, with the position amplitude for a single photon given by $\ipr{x}{\varphi_0} = E(x)/\sqrt{N}$.  The total state in the interferometer is then the tensor product $\ket{\Psi} = \ket{\psi}\ket{\varphi_0}$. In what follows, $\hbar = 1$.  

The beam will experience two distinct effects inside the interferometer that depend on the path.  First, the tilted mirror couples the system and meter by imparting a momentum kick $k$ to the transverse beam depending on the path taken, which modifies the state with the unitary operator $\op{U}_{\rm PM} = e^{ik\op{x}\op{W}}$, depending on the which-path operator.  Second, the SBC produces a net phase shift $\phi$ between the circulating modes, corresponding to the unitary operator $\op{U}_{\rm SBC} = e^{-i\phi\op{W}/2}$.  To take into account small but constant losses, we introduce the nonunitary operator $\op{L} = \sqrt{1-\gamma}\,\op{1}$, where $\gamma$ is the probability of loss per traversal from all optical imperfections.  Note that the ``loss'' of the cavity to the detector via the beamsplitter is treated separately.



After entering the interferometer through the 50:50 beamsplitter, the path state becomes an equal superposition of circulating modes $\ket{\psi_+} = \frac{1}{\sqrt{2}}\left(\ket{\cw} + i\ket{\ccw}\right)$, which is also the projection state for the bright port.  The dark port is correspondingly described by the orthogonal state $\ket{\psi_-} = \frac{1}{\sqrt{2}}\left(\ket{\cw} - i\ket{\ccw}\right)$.  Since $\op{U}_{\rm PM}$ and $\op{U}_{\rm SBC}$ are the only non-trivial actions on the system Hilbert space, it is convenient to combine their effects with the projection onto the output ports of the interferometer, which produces measurement operators $\op{M_\pm} = \bra{\psi_\pm}\op{U}_{\rm PM}\op{U}_{\rm SBC}\ket{\psi_+}$ given by
\begin{align}\label{eq:collimatedops}
  \op{M}_+ &= \cos\left(\phi/2 - k\op{x}\right), &
  \op{M}_- &= i \sin\left(\phi/2 - k\op{x}\right),
\end{align}
where $\op{x}$ is the position operator, so $\op{M}_\pm$ are diagonal in the position basis.
Here we have used $\bra{\psi_\pm}\op{W}^n\ket{\psi_+} = (1 \pm (-1)^n)/2$.  

{\it Zeno refreshing.}---After many traversals, transverse beam degradation tends to diminish the signal as discussed in \cite{Dressel2013}.  One strategy to solve this problem is to introduce a Gaussian spatial filter as shown in Fig.~\ref{figsetup} \footnote{It should be noted this is an ideal Gaussian filter (as could be well approximated by coupling into a single-mode optical fiber) rather than a typical lens-pinhole spatial filter which would suffer unacceptable diffraction and absorption losses.}.  Although a spatial filter is not essential for a successful power recycling scheme, we treat it here because it offers a more straightforward analysis compared to other arrangements.
The filter acts as a projection back onto the initial state, which can be implemented with an additional projection operator,
$\op{\Pi} = |\varphi_0\rangle \langle \varphi_0|$, so that the (normalized) state after the filter is again given by $\ket{\varphi_0}$.  The overlap between the one-pass transverse state and the initial transverse state is close to 1, indicating that the vast majority of the time, a photon will pass unimpeded through the filter.  This may be interpreted as a Zeno effect, refreshing the transverse profile, which is ideally lossless in the small $k$ limit as we will now see.
The probability of exiting the beam splitter toward the recycling mirror is given by 
\begin{equation}\label{eq:brightprob}
P_+ = |\op{M}_+ \ket{\varphi_0}|^2 = (1/2) [1+\cos \phi \exp(-2k^2 \sigma^2)],
\end{equation}
leaving the pre-filter normalized state as $\ket{\varphi'} = \op{M}_+ \ket{\varphi_0}/\sqrt{P_+}$.  The probability of surviving the filter is given by $P_Z = |\bra{\varphi_0} \varphi'\rangle|^2$,
\begin{eqnarray}
P_Z &=& (1/P_+) \left(\int dx\, \varphi_0^2(x) \cos(\phi/2 - kx)\right)^2, \nonumber \\
&=& \frac{\cos^2(\phi/2)}{\sinh(k^2 \sigma^2) + \cos^2(\phi/2) e^{- k^2 \sigma^2}}, \nonumber \\
&\approx& 1 - (\phi/2)^2 k^2 \sigma^2 -k^4 \sigma^4/2 + \ldots,
\end{eqnarray}
where the last approximation is taken in the weak value parameter range, $k \sigma \ll \phi/2 \ll 1$.  We note that the exact expression for $P_Z$ correctly equals one when $k=0$ for any value of $\phi$,
while the filter loss is $k^2\sigma^2 \phi^2/4$ to leading order. 

This small amount of loss per cycle can be incorporated into the loss from the imperfect optics (\emph{e.g.}~unwanted reflection and absorption events in the cavity) as $\gamma \rightarrow \gamma + k^2\sigma^2 \phi^2/4$, which we assume to be small compared to 1. With the filter included, the transverse beam profile is refreshed every cycle, making the calculations of the many-cycle case straightforward. The steady state amplitude exiting the detection port is given by the sum of amplitudes from all traversal numbers,
\begin{align}\label{eq:collimatedamplitude}
	\ket{\varphi} &= t \sqrt{1-\gamma}\op{M}_-  \sum_{n=0}^{\infty} \left( r\sqrt{(1-\gamma) P_+}  \right)^n \ket{\varphi_0}, \nonumber \\
	&=  \frac{i t \sqrt{1-\gamma}\sin(\phi/2 - k\op{x})}{1 - r\sqrt{(1-\gamma) P_+}}\ket{\varphi_0},
\end{align}
where $P_+ \approx \cos^2(\phi/2)$ is the probability of exiting the recycling mirror port \eqref{eq:brightprob}.  Similarly, the steady state amplitude of light reflected back towards the laser is
\begin{equation}
	\ket{\varphi_r} = \left[-r + \frac{t^2 \sqrt{1-\gamma} \cos(\phi/2)}{1 - r\sqrt{(1-\gamma) P_+}}\right]\ket{\varphi_0}, 
\end{equation}
which yields an impedance matching condition of $r = \sqrt{(1 - \gamma) P_+}$.  Notice that this choice stops any light leaking back through the recycling mirror in the steady state, regardless of the losses involved (as long as the beam remains phase-coherent, and its spectral and spatial profiles correctly realign back at the input confining mirror).  In the limit of small loss, this corresponds to setting the mirror transmission amplitude $t \approx \phi/2$.

Using an initial Gaussian profile as originally considered, the average split detector signal 
is still given by Eq.~\eqref{eq:signal}, but now
the total number of detected photons is
\begin{align}
  N_{\rm det} &= N\int_{-\infty}^{\infty}\!\! \text{d}x\, |\ipr{x}{\varphi}|^2 \approx N\,\left(1 - \frac{4\gamma}{\phi^2}\right), 
\end{align}
giving all of them, minus losses.
Therefore, the detector has an SNR of
\begin{align}\label{eq:snrresult}
  {\cal R} &\approx 4\sqrt{\frac{2}{\pi}}\,\sqrt{N}\,\frac{k\sigma}{\phi}\left[1 - \frac{2\gamma}{\phi^2}\right].  
\end{align}
From the spatial filter, the minimum loss for ideal optics is $\gamma \approx k^2\sigma^2\phi^2/4$, which produces an overall (negligible) loss factor of $(1 - k^2\sigma^2/2)$ in the SNR.
As predicted after Eq.~\eqref{eq:snr}, the SNR has been increased by the weak value factor of $(2/\phi) = 1/\sqrt{p}$ from the SNR of the single-pass weak value setup (and thus the traditional deflection setup) when the loss $\gamma \ll \phi^2/4$ is small.

{\it Power recycling with curved mirrors.}---While the recycling analysis with flat mirrors is straightforward, this cavity geometry is on the borderline between stable and unstable \cite{Milonni2010}.
In practice, the beam will be a diffracting Gaussian beam with a curved phase front rather than the collimated beam treated above.  To confine such a Gaussian beam in a stable configuration, the radius of curvature of the recycling mirror must match the radius of curvature of the phase fronts to ensure proper phase cancellation at the mirror. This geometry is sketched in Fig.~\ref{figsetup}(b). The remaining flat mirrors in our setup have no effect on the confinement properties. The resonant cavity is characterized by its {\it finesse}, the typical number of bounces before the beam decays.  In order to maximize the gains from the recycling procedure, the finesse should exceed the inverse probability to exit the interferometer to the detector.
Placing the beam focus at the far mirror ensures that the symmetry of the interferometer paths is not disturbed by the changing beam waist, enabling proper interference of the clockwise and counter-clockwise propagating curved phase fronts at the beam splitter.  This geometry defines a symmetric confocal cavity, which has well-established properties.  The confocal cavity lies at the other stability extreme of the plane parallel geometry \cite{Milonni2010}.

The physics of this cavity is similar to that of the collimated analysis, with a few important differences.  The beam will achieve its minimum waist $\sigma_0$ at the far mirror and its maximum waist $\sigma$ at the recycling mirror, where it should match the spatial and spectral profiles of the input beam.  If the coordinate $z$ along the optical axis is measured from the minimum waist at the symmetry point, and the maximum cavity length between that point and recycling mirror is $\ell$, then the beam width inside the cavity is given by $\sigma(z) = \sigma_0\sqrt{1 + z^2/\ell^2}$, where $\ell$ is also equal to the Rayleigh range, and $\sigma(\ell) = \sqrt{2}\,\sigma_0 \equiv \sigma$ is the input beam waist. 
Putting the transverse mirror momentum kick $k$ at the focus of the cavity does not yield a sensitive response, so we put the momentum kick instead on the beam splitter, as was done in the experiments of Ref.~\cite{Viza2014}.  The transmitted beams acquire no momentum kick, while the reflected beams acquire a momentum kick $k$.
The presence of a focus in the cavity gives two new effects.  The first is that a Gouy phase appears from the focus, giving an additional phase factor of $\pi$ in both beams \cite{Milonni2010}.  The second more important effect is that passing through the focus flips the tilt of the phase front, so the effective transverse momentum kick from the beam splitter is inverted when the expanding beam returns to the beam splitter, $k \rightarrow -k$.  If left uncorrected, this momentum kick is undone by the additional $k$ momentum kick from the second reflection to the detector.  We can compensate for this effect by adding a Dove prism inside the interferometer, which provides a transverse parity flip to restore the previous phase front, recovering similar weak value physics to the collimated case.  The only significant difference from the previous analysis 
concerns the changing width of the beam $\sigma(z)$.  The choice of cavity geometry will set the width $\sigma$ in Eq.~(\ref{eq:snrresult}).

{\it Conclusion.}---By including a power-recycling mirror in a continuous wave interferometric weak value amplification setup, we are able to maintain the large pointer shift associated with previous weak value amplification experiments while acquiring all of the input light in principle.  Our main result is that the SNR (or, equivalently, the Fisher information about the desired parameter) is boosted by the weak value factor, which can be made large in principle, limited only by the fidelity of the optics and the finesse of the cavity.  We have given two different cavity geometries to realize this proposal, but other stable geometries giving similar physics also exist.

In this work, we have focused on the interferometric implementation of the optical weak value effect to propose the use of the power recycling technique.  However, the same basic idea may be applied to other experimental realizations of the same, such as the polarization-based version \cite{Hosten2008}, where the postselection is accomplished with a polarizing beam splitter, and the other output beam is reinjected into the experiment.

Power recycling is only one of the techniques used in precision interferometric measurements. There are several others which may be able to be combined with our setup as well.  As for further improvements in sensitivity, we have already discussed the possibility of recycling the signal.  Future work may focus on the combination of this technique and quantum light metrological approaches such as using squeezed and entangled states \cite{Caves1981,Treps2002}.

{\it Acknowledgments.}---This work was supported by the Army Research Office grant number W911NF-12-1-0263 and W911NF-09-0-01417, as well as the CAPES Foundation, process number BEX 8257/13-2. It was also partially funded by the ARO MURI Grant No. W911NF-11-1-0268, as well as the ODNI-IARPA-ARO Grant No. W911NF-10-1-0334. We thank A. Steinberg and T. Brun for comments on this work.

\bibliography{main}

\begin{thebibliography}{19}%
\makeatletter
\providecommand \@ifxundefined [1]{%
 \@ifx{#1\undefined}
}%
\providecommand \@ifnum [1]{%
 \ifnum #1\expandafter \@firstoftwo
 \else \expandafter \@secondoftwo
 \fi
}%
\providecommand \@ifx [1]{%
 \ifx #1\expandafter \@firstoftwo
 \else \expandafter \@secondoftwo
 \fi
}%
\providecommand \natexlab [1]{#1}%
\providecommand \enquote  [1]{``#1''}%
\providecommand \bibnamefont  [1]{#1}%
\providecommand \bibfnamefont [1]{#1}%
\providecommand \citenamefont [1]{#1}%
\providecommand \href@noop [0]{\@secondoftwo}%
\providecommand \href [0]{\begingroup \@sanitize@url \@href}%
\providecommand \@href[1]{\@@startlink{#1}\@@href}%
\providecommand \@@href[1]{\endgroup#1\@@endlink}%
\providecommand \@sanitize@url [0]{\catcode `\\12\catcode `\$12\catcode
  `\&12\catcode `\#12\catcode `\^12\catcode `\_12\catcode `\%12\relax}%
\providecommand \@@startlink[1]{}%
\providecommand \@@endlink[0]{}%
\providecommand \url  [0]{\begingroup\@sanitize@url \@url }%
\providecommand \@url [1]{\endgroup\@href {#1}{\urlprefix }}%
\providecommand \urlprefix  [0]{URL }%
\providecommand \Eprint [0]{\href }%
\providecommand \doibase [0]{http://dx.doi.org/}%
\providecommand \selectlanguage [0]{\@gobble}%
\providecommand \bibinfo  [0]{\@secondoftwo}%
\providecommand \bibfield  [0]{\@secondoftwo}%
\providecommand \translation [1]{[#1]}%
\providecommand \BibitemOpen [0]{}%
\providecommand \bibitemStop [0]{}%
\providecommand \bibitemNoStop [0]{.\EOS\space}%
\providecommand \EOS [0]{\spacefactor3000\relax}%
\providecommand \BibitemShut  [1]{\csname bibitem#1\endcsname}%
\let\auto@bib@innerbib\@empty
\bibitem [{\citenamefont {Aharonov}\ \emph {et~al.}(1988)\citenamefont
  {Aharonov}, \citenamefont {Albert},\ and\ \citenamefont
  {Vaidman}}]{Aharonov1988}%
  \BibitemOpen
  \bibfield  {author} {\bibinfo {author} {\bibfnamefont {Y.}~\bibnamefont
  {Aharonov}}, \bibinfo {author} {\bibfnamefont {D.~Z.}\ \bibnamefont
  {Albert}}, \ and\ \bibinfo {author} {\bibfnamefont {L.}~\bibnamefont
  {Vaidman}},\ }\href@noop {} {\bibfield  {journal} {\bibinfo  {journal} {Phys.
  Rev. Lett.}\ }\textbf {\bibinfo {volume} {60}},\ \bibinfo {pages} {1351 }
  (\bibinfo {year} {1988})}\BibitemShut {NoStop}%
\bibitem [{\citenamefont {Starling}\ \emph {et~al.}(2010)\citenamefont
  {Starling}, \citenamefont {Dixon}, \citenamefont {Jordan},\ and\
  \citenamefont {Howell}}]{Starling2010}%
  \BibitemOpen
  \bibfield  {author} {\bibinfo {author} {\bibfnamefont {D.~J.}\ \bibnamefont
  {Starling}}, \bibinfo {author} {\bibfnamefont {P.~B.}\ \bibnamefont {Dixon}},
  \bibinfo {author} {\bibfnamefont {A.~N.}\ \bibnamefont {Jordan}}, \ and\
  \bibinfo {author} {\bibfnamefont {J.~C.}\ \bibnamefont {Howell}},\
  }\href@noop {} {\bibfield  {journal} {\bibinfo  {journal} {Phys. Rev. A}\
  }\textbf {\bibinfo {volume} {82}},\ \bibinfo {pages} {063822} (\bibinfo
  {year} {2010})}\BibitemShut {NoStop}%
\bibitem [{\citenamefont {Jordan}\ \emph {et~al.}(2014)\citenamefont {Jordan},
  \citenamefont {Mart\'inez-Rinc\'on},\ and\ \citenamefont
  {Howell}}]{Jordan2014}%
  \BibitemOpen
  \bibfield  {author} {\bibinfo {author} {\bibfnamefont {A.~N.}\ \bibnamefont
  {Jordan}}, \bibinfo {author} {\bibfnamefont {J.}~\bibnamefont
  {Mart\'inez-Rinc\'on}}, \ and\ \bibinfo {author} {\bibfnamefont {J.~C.}\
  \bibnamefont {Howell}},\ }\href@noop {} {\bibfield  {journal} {\bibinfo
  {journal} {Phys. Rev. X}\ }\textbf {\bibinfo {volume} {4}},\ \bibinfo {pages}
  {011031} (\bibinfo {year} {2014})}\BibitemShut {NoStop}%
\bibitem [{\citenamefont {Viza}\ \emph {et~al.}(2014)\citenamefont {Viza},
  \citenamefont {Mart\'inez-Rinc\'on}, \citenamefont {Alves}, \citenamefont
  {Jordan},\ and\ \citenamefont {Howell}}]{Viza2014}%
  \BibitemOpen
  \bibfield  {author} {\bibinfo {author} {\bibfnamefont {G.~I.}\ \bibnamefont
  {Viza}}, \bibinfo {author} {\bibfnamefont {J.}~\bibnamefont
  {Mart\'inez-Rinc\'on}}, \bibinfo {author} {\bibfnamefont {G.~B.}\
  \bibnamefont {Alves}}, \bibinfo {author} {\bibfnamefont {A.~N.}\ \bibnamefont
  {Jordan}}, \ and\ \bibinfo {author} {\bibfnamefont {J.~C.}\ \bibnamefont
  {Howell}},\ }\href@noop {} {\bibfield  {journal} {\bibinfo  {journal}
  {arXiv:1410.8461}\ } (\bibinfo {year} {2014})}\BibitemShut {NoStop}%
\bibitem [{\citenamefont {Pang}\ \emph {et~al.}(2014)\citenamefont {Pang},
  \citenamefont {Dressel},\ and\ \citenamefont {Brun}}]{Pang2014}%
  \BibitemOpen
  \bibfield  {author} {\bibinfo {author} {\bibfnamefont {S.}~\bibnamefont
  {Pang}}, \bibinfo {author} {\bibfnamefont {J.}~\bibnamefont {Dressel}}, \
  and\ \bibinfo {author} {\bibfnamefont {T.~A.}\ \bibnamefont {Brun}},\
  }\href@noop {} {\bibfield  {journal} {\bibinfo  {journal} {Phys. Rev. Lett.}\
  }\textbf {\bibinfo {volume} {113}},\ \bibinfo {pages} {030401} (\bibinfo
  {year} {2014})}\BibitemShut {NoStop}%
\bibitem [{\citenamefont {Hosten}\ and\ \citenamefont
  {Kwiat}(2008)}]{Hosten2008}%
  \BibitemOpen
  \bibfield  {author} {\bibinfo {author} {\bibfnamefont {O.}~\bibnamefont
  {Hosten}}\ and\ \bibinfo {author} {\bibfnamefont {P.}~\bibnamefont {Kwiat}},\
  }\href@noop {} {\bibfield  {journal} {\bibinfo  {journal} {Science}\ }\textbf
  {\bibinfo {volume} {319}},\ \bibinfo {pages} {787} (\bibinfo {year}
  {2008})}\BibitemShut {NoStop}%
\bibitem [{\citenamefont {Dixon}\ \emph {et~al.}(2009)\citenamefont {Dixon},
  \citenamefont {Starling}, \citenamefont {Jordan},\ and\ \citenamefont
  {Howell}}]{Dixon2009}%
  \BibitemOpen
  \bibfield  {author} {\bibinfo {author} {\bibfnamefont {P.~B.}\ \bibnamefont
  {Dixon}}, \bibinfo {author} {\bibfnamefont {D.~J.}\ \bibnamefont {Starling}},
  \bibinfo {author} {\bibfnamefont {A.~N.}\ \bibnamefont {Jordan}}, \ and\
  \bibinfo {author} {\bibfnamefont {J.~C.}\ \bibnamefont {Howell}},\
  }\href@noop {} {\bibfield  {journal} {\bibinfo  {journal} {Phys. Rev. Lett.}\
  }\textbf {\bibinfo {volume} {102}},\ \bibinfo {pages} {173601} (\bibinfo
  {year} {2009})}\BibitemShut {NoStop}%
\bibitem [{\citenamefont {Hogan}\ \emph {et~al.}(2011)\citenamefont {Hogan},
  \citenamefont {Hammer}, \citenamefont {Chiow}, \citenamefont {Dickerson},
  \citenamefont {Johnson}, \citenamefont {Kovachy}, \citenamefont
  {Sugerbaker},\ and\ \citenamefont {Kasevich}}]{Hogan2011}%
  \BibitemOpen
  \bibfield  {author} {\bibinfo {author} {\bibfnamefont {H.}~\bibnamefont
  {Hogan}}, \bibinfo {author} {\bibfnamefont {J.}~\bibnamefont {Hammer}},
  \bibinfo {author} {\bibfnamefont {S.-W.}\ \bibnamefont {Chiow}}, \bibinfo
  {author} {\bibfnamefont {S.}~\bibnamefont {Dickerson}}, \bibinfo {author}
  {\bibfnamefont {D.~M.~S.}\ \bibnamefont {Johnson}}, \bibinfo {author}
  {\bibfnamefont {T.}~\bibnamefont {Kovachy}}, \bibinfo {author} {\bibfnamefont
  {A.}~\bibnamefont {Sugerbaker}}, \ and\ \bibinfo {author} {\bibfnamefont
  {M.~A.}\ \bibnamefont {Kasevich}},\ }\href@noop {} {\bibfield  {journal}
  {\bibinfo  {journal} {Optics letters}\ }\textbf {\bibinfo {volume} {36}},\
  \bibinfo {pages} {1698} (\bibinfo {year} {2011})}\BibitemShut {NoStop}%
\bibitem [{\citenamefont {Kofman}\ \emph {et~al.}(2012)\citenamefont {Kofman},
  \citenamefont {Ashkab},\ and\ \citenamefont {Nori}}]{Kofman2011}%
  \BibitemOpen
  \bibfield  {author} {\bibinfo {author} {\bibfnamefont {A.~G.}\ \bibnamefont
  {Kofman}}, \bibinfo {author} {\bibfnamefont {S.}~\bibnamefont {Ashkab}}, \
  and\ \bibinfo {author} {\bibfnamefont {F.}~\bibnamefont {Nori}},\ }\href@noop
  {} {\bibfield  {journal} {\bibinfo  {journal} {Physics Reports}\ }\textbf
  {\bibinfo {volume} {520}},\ \bibinfo {pages} {43} (\bibinfo {year}
  {2012})}\BibitemShut {NoStop}%
\bibitem [{\citenamefont {Dressel}\ \emph {et~al.}(2014)\citenamefont
  {Dressel}, \citenamefont {Malik}, \citenamefont {Miatto}, \citenamefont
  {Jordan},\ and\ \citenamefont {Boyd}}]{Dressel2014}%
  \BibitemOpen
  \bibfield  {author} {\bibinfo {author} {\bibfnamefont {J.}~\bibnamefont
  {Dressel}}, \bibinfo {author} {\bibfnamefont {M.}~\bibnamefont {Malik}},
  \bibinfo {author} {\bibfnamefont {F.~M.}\ \bibnamefont {Miatto}}, \bibinfo
  {author} {\bibfnamefont {A.~N.}\ \bibnamefont {Jordan}}, \ and\ \bibinfo
  {author} {\bibfnamefont {R.~W.}\ \bibnamefont {Boyd}},\ }\href@noop {}
  {\bibfield  {journal} {\bibinfo  {journal} {Rev. Mod. Phys.}\ }\textbf
  {\bibinfo {volume} {86}},\ \bibinfo {pages} {307} (\bibinfo {year}
  {2014})}\BibitemShut {NoStop}%
\bibitem [{\citenamefont {Drever}(1983)}]{Drever1983}%
  \BibitemOpen
  \bibfield  {author} {\bibinfo {author} {\bibfnamefont {R.}~\bibnamefont
  {Drever}},\ }in\ \href@noop {} {\emph {\bibinfo {booktitle} {Quantum Optics:
  Experimental Gravity and Measurement Theory}}}\ (\bibinfo  {publisher}
  {Plenum, New York},\ \bibinfo {year} {1983})\ pp.\ \bibinfo {pages}
  {525--566}\BibitemShut {NoStop}%
\bibitem [{\citenamefont {Dressel}\ \emph {et~al.}(2013)\citenamefont
  {Dressel}, \citenamefont {Lyons}, \citenamefont {Jordan}, \citenamefont
  {Graham},\ and\ \citenamefont {Kwiat}}]{Dressel2013}%
  \BibitemOpen
  \bibfield  {author} {\bibinfo {author} {\bibfnamefont {J.}~\bibnamefont
  {Dressel}}, \bibinfo {author} {\bibfnamefont {K.}~\bibnamefont {Lyons}},
  \bibinfo {author} {\bibfnamefont {A.~N.}\ \bibnamefont {Jordan}}, \bibinfo
  {author} {\bibfnamefont {T.}~\bibnamefont {Graham}}, \ and\ \bibinfo {author}
  {\bibfnamefont {P.}~\bibnamefont {Kwiat}},\ }\href@noop {} {\bibfield
  {journal} {\bibinfo  {journal} {Phys. Rev. A}\ }\textbf {\bibinfo {volume}
  {88}},\ \bibinfo {pages} {023821} (\bibinfo {year} {2013})}\BibitemShut
  {NoStop}%
\bibitem [{\citenamefont {Starling}\ \emph {et~al.}(2009)\citenamefont
  {Starling}, \citenamefont {Dixon}, \citenamefont {Jordan},\ and\
  \citenamefont {Howell}}]{Starling2009}%
  \BibitemOpen
  \bibfield  {author} {\bibinfo {author} {\bibfnamefont {D.~J.}\ \bibnamefont
  {Starling}}, \bibinfo {author} {\bibfnamefont {P.~B.}\ \bibnamefont {Dixon}},
  \bibinfo {author} {\bibfnamefont {A.~N.}\ \bibnamefont {Jordan}}, \ and\
  \bibinfo {author} {\bibfnamefont {J.~C.}\ \bibnamefont {Howell}},\
  }\href@noop {} {\bibfield  {journal} {\bibinfo  {journal} {Phys. Rev. A}\
  }\textbf {\bibinfo {volume} {80}},\ \bibinfo {pages} {041803} (\bibinfo
  {year} {2009})}\BibitemShut {NoStop}%
\bibitem [{\citenamefont {Schnier}\ \emph {et~al.}(1997)\citenamefont
  {Schnier}, \citenamefont {Mizuno}, \citenamefont {Heinzel}, \citenamefont
  {L\"{u}ck}, \citenamefont {R\"udinger}, \citenamefont {Schilling},
  \citenamefont {Schrempel}, \citenamefont {Winkler},\ and\ \citenamefont
  {Danzmann}}]{Schnier1997}%
  \BibitemOpen
  \bibfield  {author} {\bibinfo {author} {\bibfnamefont {D.}~\bibnamefont
  {Schnier}}, \bibinfo {author} {\bibfnamefont {J.}~\bibnamefont {Mizuno}},
  \bibinfo {author} {\bibfnamefont {G.}~\bibnamefont {Heinzel}}, \bibinfo
  {author} {\bibfnamefont {H.}~\bibnamefont {L\"{u}ck}}, \bibinfo {author}
  {\bibfnamefont {A.}~\bibnamefont {R\"udinger}}, \bibinfo {author}
  {\bibfnamefont {R.}~\bibnamefont {Schilling}}, \bibinfo {author}
  {\bibfnamefont {M.}~\bibnamefont {Schrempel}}, \bibinfo {author}
  {\bibfnamefont {W.}~\bibnamefont {Winkler}}, \ and\ \bibinfo {author}
  {\bibfnamefont {K.}~\bibnamefont {Danzmann}},\ }\href@noop {} {\bibfield
  {journal} {\bibinfo  {journal} {Phys. Lett. A}\ }\textbf {\bibinfo {volume}
  {225}},\ \bibinfo {pages} {210} (\bibinfo {year} {1997})}\BibitemShut
  {NoStop}%
\bibitem [{\citenamefont {Meers}\ and\ \citenamefont
  {Strain}(1991)}]{Meers1991}%
  \BibitemOpen
  \bibfield  {author} {\bibinfo {author} {\bibfnamefont {B.}~\bibnamefont
  {Meers}}\ and\ \bibinfo {author} {\bibfnamefont {K.}~\bibnamefont {Strain}},\
  }\href@noop {} {\bibfield  {journal} {\bibinfo  {journal} {Phys. Rev. D}\
  }\textbf {\bibinfo {volume} {43}},\ \bibinfo {pages} {3117} (\bibinfo {year}
  {1991})}\BibitemShut {NoStop}%
\bibitem [{Note1()}]{Note1}%
  \BibitemOpen
  \bibinfo {note} {It should be noted this is an ideal Gaussian filter (as
  could be well approximated by coupling into a single-mode optical fiber)
  rather than a typical lens-pinhole spatial filter which would suffer
  unacceptable diffraction and absorption losses.}\BibitemShut {Stop}%
\bibitem [{\citenamefont {Milonni}\ and\ \citenamefont
  {Eberly}(2010)}]{Milonni2010}%
  \BibitemOpen
  \bibfield  {author} {\bibinfo {author} {\bibfnamefont {P.~W.}\ \bibnamefont
  {Milonni}}\ and\ \bibinfo {author} {\bibfnamefont {J.~H.}\ \bibnamefont
  {Eberly}},\ }\href@noop {} {\emph {\bibinfo {title} {Laser Physics}}}\
  (\bibinfo  {publisher} {Wiley},\ \bibinfo {year} {2010})\ pp.\ \bibinfo
  {pages} {269--327}\BibitemShut {NoStop}%
\bibitem [{\citenamefont {Caves}(1981)}]{Caves1981}%
  \BibitemOpen
  \bibfield  {author} {\bibinfo {author} {\bibfnamefont {C.~M.}\ \bibnamefont
  {Caves}},\ }\href@noop {} {\bibfield  {journal} {\bibinfo  {journal} {Phys.
  Rev. D}\ }\textbf {\bibinfo {volume} {23}},\ \bibinfo {pages} {1693}
  (\bibinfo {year} {1981})}\BibitemShut {NoStop}%
\bibitem [{\citenamefont {Treps}\ \emph {et~al.}(2002)\citenamefont {Treps},
  \citenamefont {Andersen}, \citenamefont {Bulcher}, \citenamefont {Lam},
  \citenamefont {Maitre}, \citenamefont {Bachor},\ and\ \citenamefont
  {Fabre}}]{Treps2002}%
  \BibitemOpen
  \bibfield  {author} {\bibinfo {author} {\bibfnamefont {N.}~\bibnamefont
  {Treps}}, \bibinfo {author} {\bibfnamefont {U.}~\bibnamefont {Andersen}},
  \bibinfo {author} {\bibfnamefont {B.}~\bibnamefont {Bulcher}}, \bibinfo
  {author} {\bibfnamefont {P.~K.}\ \bibnamefont {Lam}}, \bibinfo {author}
  {\bibfnamefont {A.}~\bibnamefont {Maitre}}, \bibinfo {author} {\bibfnamefont
  {H.-A.}\ \bibnamefont {Bachor}}, \ and\ \bibinfo {author} {\bibfnamefont
  {C.}~\bibnamefont {Fabre}},\ }\href@noop {} {\bibfield  {journal} {\bibinfo
  {journal} {Phys. Rev. Lett.}\ }\textbf {\bibinfo {volume} {88}},\ \bibinfo
  {pages} {203601} (\bibinfo {year} {2002})}\BibitemShut {NoStop}%
\end{thebibliography}%

\end{document}